\documentclass{aa}
\usepackage{txfonts}
\usepackage{graphicx}

\begin{document}
\title{Non-symmetric magnetohydrostatic equilibria: a multigrid approach}

\author{D. MacTaggart\inst{1}, A. Elsheikh\inst{1}, J.~A.~McLaughlin\inst{2} \and R.~D.~Simitev\inst{3}}

\offprints{D. MacTaggart, \email{D.MacTaggart@abertay.ac.uk}}

\institute{School of Engineering, Computing and Applied Mathematics, University of Abertay Dundee, Kydd Building, Dundee, DD1 1HG, Scotland, UK \and Department of Mathematics and Information Sciences, Northumbria University, Newcastle Upon Tyne, NE1 8ST, England, UK \and School of Mathematics and Statistics, University of Glasgow, Glasgow, G12 8QW, Scotland, UK}

\date{}

\abstract {} {Linear magnetohydrostatic (MHS) models of solar magnetic fields balance plasma pressure gradients, gravity and Lorentz forces where the current density is composed of a linear force-free component and a cross-field component that depends on gravitational stratification. In this paper, we investigate an efficient numerical procedure for calculating such equilibria.}{The MHS equations are reduced to two scalar elliptic equations - one on the lower boundary and the other within the interior of the computational domain. The normal component of the magnetic field is prescribed on the lower boundary and a multigrid method is applied on both this boundary and within the domain to find the poloidal scalar potential. Once solved to a desired accuracy, the magnetic field, plasma pressure and density are found using a finite difference method.}{We investigate the effects of the cross-field currents on the linear MHS equilibria. Force-free and non-force-free examples are given to demonstrate the numerical scheme and an analysis of speed-up due to parallelization on a graphics processing unit (GPU) is presented. It is shown that speed-ups of $\times$30 are readily achievable.} {}


\titlerunning{Non-symmetric MHS equilibria: a multigrid approach}
\authorrunning{MacTaggart et al.}
\maketitle

\section{Introduction}
The calculation of three-dimensional non-linear magnetohydrostatic (MHS) equilibria is a non-trivial subject. In the solar physics context, progress has been made by considering linear subclasses of MHS equilibria. One such class, known as laminated equilibria, makes use of an Euler potential representation of the magnetic field (e.g. \cite{low82}). On each lamina, a 2D magnetic field is calculated and the 3D field comprises of the union of these laminas. 

Another approach is to model the current density as a linear combination of field-aligned and cross-field currents (\cite{low91,low92}). Here, the field-aligned current is that of a linear force-free field and the cross-field current depends on the variation of the magnetic field with height. Details of this will be presented in the next section. \cite{neukirch99} (NR99 hereafter) present a new formulation of Low's model by writing the magnetic field in terms of poloidal and toroidal components. This has the advantage that the calculation of the magnetic field only involves one scalar function whereas, previously, one was forced to operate with all three components of the magnetic field
independently. \cite{petrie00} use the representation of NR99 and find closed-form solutions for MHS equilibria via Green's functions. The price paid for finding closed-form solutions, however, is that they are forced to choose a simple term for the cross-field current. To date, the authors are unaware of any closed-form solutions using Green's functions, other than those presented in \cite{petrie00}.

Although more detailed analytical solutions may prove difficult with the representation of NR99, it is, however, set up perfectly for an efficient numerical treatment. In this paper, we outline a simple and fast numerical procedure for calculating MHS equilibria based on the NR99 representation. The details of this are given in the next section. This is followed by some examples of force-free and linear MHS equilibria. For the linear MHS case, we investigate the effects of the cross-field currents on the equilibria. We then highlight how the scheme can be parallelized simply and effectively on a graphics processing unit (GPU).  The paper concludes with a summary.

\section{Model equations and solution method}
Firstly, we shall outline where the equations to be solved come from. Fuller details can be found in NR99. This will be followed by an algorithm for the numerical solution. 

\subsection{The model equations}
The MHS equations that we shall solve are
\begin{equation}\label{mhs1}
\mu_0^{-1}(\nabla\times\mathbf{B})\times\mathbf{B} -\nabla p - \rho g\hat{\mathbf{e}}_z = \mathbf{0},
\end{equation}

\begin{equation}\label{mhs2}
\nabla\times\mathbf{B} = \mu_0\mathbf{j},
\end{equation}

\begin{equation}\label{mhs3}
\nabla\cdot\mathbf{B} = 0,
\end{equation}
where $\mathbf{B}$ is the magnetic induction (commonly referred to as the magnetic field), $\mathbf{j}$ is the current density, $p$ is the plasma presure, $\rho$ is the density, $g$ is the (constant) gravitational acceleration and $\mu_0$ is the permeability of free space. To complete the problem, boundary conditions must be specified. This choice is problem-dependent but will, at least, require $\mathbf{B}$ to be specified on the boundaries of the domain through Dirichlet or von Neuman conditions. In this paper we will consider a Cartesian domain. Following \cite{low92}, we assume the current density takes the following form
\begin{equation}\label{current}
\mu_0\mathbf{j} = \alpha\mathbf{B} + \nabla(gF)\times\hat{\mathbf{e}}_z,
\end{equation}
where $\alpha$ is a constant and $F$ is an arbitrary function. The first term on the right-hand-side is the aligned current associated with a linear force-free field. The second term is the non-force-free cross current due to the gravitational stratification. The magnetic induction can be written as 
\begin{equation}\label{b_iden}
\mathbf{B} = \nabla\times[\nabla\times(P\hat{\mathbf{e}}_z) + T\hat{\mathbf{e}}_z],
\end{equation}
where $P$ and $T$ are scalar functions corresponding to the poloidal and toroidal components respectively. This form satisfies equation (\ref{mhs3}). Using this in equation (\ref{mhs2}) with equation (\ref{current}) gives
\begin{equation}\label{t}
T = \alpha P,
\end{equation}

\begin{equation}\label{helm}
\nabla^2P + \alpha^2 P + gF = 0.
\end{equation}
Following NR99 and \cite{low92}, we set
\[
F = g^{-1}\xi(z)B_z.
\]
With this identity, equation (\ref{helm}) becomes
\begin{equation}\label{scalar_pot}
[1-\xi(z)]\left(\frac{\partial^2P}{\partial x^2} +\frac{\partial^2P}{\partial y^2}\right) + \frac{\partial^2P}{\partial z^2} + \alpha^2 P = 0.
\end{equation}
For $\xi(z) < 1$, equation (\ref{scalar_pot}) is elliptic. Now the magnetic field can be found by solving this for the poloidal scalar potential $P$ and then using equations (\ref{b_iden}) and (\ref{t}).

Once the magnetic field is found, the plasma pressure and density are given by
\begin{equation}\label{pressure}
p = p_b(z) - \xi(z)\frac{B_z^2}{2\mu_0}
\end{equation}
and
\begin{equation}\label{density}
\rho = \rho_b(z) + \frac{1}{g}\left(\frac{{\rm d}\xi}{{\rm d}z}\frac{B_z^2}{2\mu_0} + \frac{1}{\mu_0}\xi~\mathbf{B}\cdot\nabla B_z \right).
\end{equation}
$p_b$ and $\rho_b$ are the background equilibrium plasma pressure and density respectively. A derivation of these formulae is given in NR99.

\subsection{The numerical procedure}
An efficient numerical solution of this class of the MHS equations rests on reducing the equations to a set of linear, scalar elliptic problems. i.e. those of the form
\[
a(\mathbf{x})\nabla^2u +b(\mathbf{x})u =c(\mathbf{x}), \qquad \mathbf{x}\in \mathcal{D},
\]
with $u$ defined on $\partial\mathcal{D}$. Here the Laplacian operator can be two-dimensional (for boundaries) or three-dimensional (for the interior) and $a$, $b$ and $c$ are known scalar functions.
 
There exists a wide variety of techniques to solve such problems efficiently, from Krylov subspace methods to multigrid methods. Although solution methods are problem-dependent, the multigrid method is often an optimal solver for discrete Poisson problems (\cite{elman05}), i.e. its convergence rate is independent of the problem mesh size. It can also be parallelized efficiently, and for these reasons we adopt it in this work. The multigrid method, in a different representation to the one of this paper, was used successfully for MHS models of flux tubes bounded by current sheets (\cite{henning01}).

As mentioned above, the MHS equations are reduced to a set of linear elliptic problems defined on the boundaries and in the interior. Equation (\ref{scalar_pot}) is the three-dimensional elliptic problem for the poloidal scalar potential $P$ defined in the interior of the domain. The other elliptic equations required to be solved depend on the boundary conditions. On the lower boundary, the vertical component of the magnetic field, $B_z$, is prescribed. From equation (\ref{b_iden}), it is clear that, on the lower boundary,
\begin{equation}\label{lower_bound}
\left(\frac{\partial^2P}{\partial x^2} +\frac{\partial^2P}{\partial y^2}\right)=-B_z(x,y).
\end{equation}
In principle other elliptic equtions can be defined on the other boundaries of the Cartesian domain. For simplicity, however, we shall assume the $\mathbf{B}=\mathbf{0}$ on these boundaries. For the scalar elliptic problem, this translates to $P=0$ on top and side boundaries. This boundary condition also applies to the sides of the 2D domain (the lower boundary) where equation (\ref{lower_bound}) is solved.
In short, the numerical procedure to determine the poloidal scalar potential $P$ involves:
\begin{enumerate}
\item Set $P=0$ everywhere, initially, on the grid.
\item Use the multigrid method to solve equation (\ref{lower_bound}) on the lower boundary with $P=0$ on the sides.
\item Use the multigrid method to solve equation (\ref{scalar_pot}) within the domain, using $P=0$ on the top and side boundaries and the distribution of $P$ on the lower boundary determined from the previous step.
\end{enumerate}
After these steps are completed, the magnetic field $\mathbf{B}$ can be calculated. From equation (\ref{b_iden}), the other components of the magnetic field are given by
\[
B_x = \alpha\frac{\partial P}{\partial y} + \frac{\partial^2 P}{\partial x\partial z}
\]
and
\[
B_y=-\alpha\frac{\partial P}{\partial x} + \frac{\partial^2 P}{\partial y\partial z}.
\]
These, together with equation (\ref{lower_bound}), can be approximated using finite differences, giving $B_x$, $B_y$ and $B_z$ on the grid to a required accuracy. In this paper, we use second-order accurate central finite differences.

All that remains now is the calculation of the pressure $p$ and the density $\rho$. Once $B_z$ is determined on the grid, it is clear from equation (\ref{pressure}) that $p$ can be evaluated directly. In equation (\ref{density}) the derivatives must be dealt with using finite differences. After this, $\rho$ is determined on the grid.

\section{Examples} 
Here we present examples of equilibria to demonstrate the effectiveness of our numerical scheme. Details of computational aspects and parallelization are discussed in Section 4. 

For the cases under consideration we non-dimensionalize equations (\ref{mhs1}) to (\ref{mhs3}) with respect to photospheric values. The variables are plasma pressure, $p_0=1.4\times 10^4$ Pa; density, $\rho_0 = 3\times 10^{-4}$ kg/m$^3$; scale height $H_0=340$ km; magnetic induction $B_0 = (2\mu_0p_0)^{1/2}=1.3\times 10^3$ G and temperature, $T_0 = p_0/(R\rho_0) = 5.6\times 10^3$ K. Here, $R$ is the gas constant.

We assume a background temperature profile of the form
\[
T_b(z) = \left\{\begin{array}{cc}
1, & 0\le z\le 5,  \\
T_{\rm cor}^{(z-5)/5}, & 5<z\le 10, \\
T_{\rm cor}, & z > 10.\\
\end{array}\right.
\]
$T_{\rm cor}$=150 is the non-dimensional coronal temperature. The model photosphere/chromosphere ranges from $0\le z\le 5$, the transition region ranges from $5<z\le 10$ and the corona is in $z > 10$. With an ideal gas equation
\[
p_b=\rho_b T_b,
\]
hydrostatic pressure balance can be written as
\[
\frac{{\rm d}p_b}{{\rm d}z} = -\frac{p_bg}{T_b}.
\]
The equation is solved for $p_b$ and then $\rho_b$ follows from the ideal gas law. The non-dimensionalization and atmospheric model presented here are similar to those in flux emergence studies (e.g. \cite{dmac11,hood12,mclaughlin12}). For the following cases, we set $\mathbf{B}=\mathbf{0}$ on the side and top boundaries. The size of the domain is $[-5.5,5.5]^2\times[0,12]$ and we use a resolution of 128$^3$. 
\subsection{Force-free case}
\begin{figure}
 \resizebox{\hsize}{!}{\includegraphics{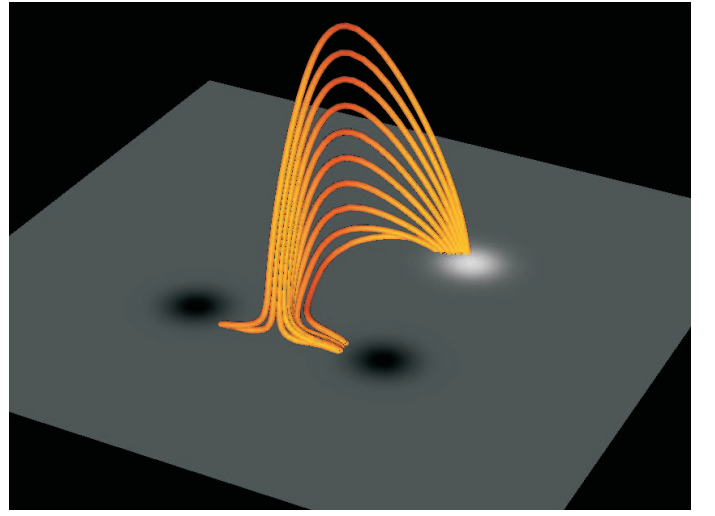}}
 \caption{Magnetic field lines of the linear force-free region are displayed in orange. A magnetogram of $B_z$ is given at the base to indicate the positions of the sources.}
\label{lfff}
\end{figure}

For a force-free field, $\xi=0$. This, of course, means that $p=p_b$ and $\rho=\rho_b$. In this example we take $\alpha=0.4$. We model a magnetic configuration with three sources, similar to that in \cite{regnier05}. For each source, we define a Gaussian profile for $B_z$, on the lower boundary, of the form
\[
B_z(r) = B_0\exp(-{r^2}/{l^2}).
\]
$B_0$ is the field strength at the centre of the source, $l$ is source width and $r^2=(x-x_0)^2+(y-y_0)^2$ with source centre $(x_0,y_0)$. Here we take $l=0.3$ for all the sources. There is one positive source with $B_z=1$ and two negative sources with $B_z=-0.5$. The position of the positive source is (1.5,1.5). The two negative sources have positions (-1.5,-1.5) and (1.5,-1.5). Figure \ref{lfff} shows some field lines of the calculated region, with a magnetogram of $B_z$ displayed at the base.

\begin{figure}
 \resizebox{\hsize}{!}{\includegraphics{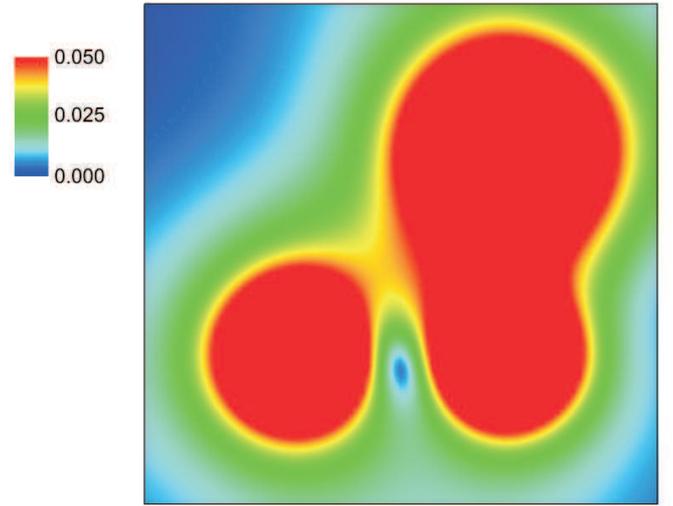}}
 \caption{A map of $|\mathbf{B}|$, within the range given on the scale, at $z=0.1$. The null point is clearly highlighted between the two negative polarities.}
\label{bmag}
\end{figure}

Between the two negative polarities, there is a rapid divergence in the field lines as they connect down to the positive polarity. This indicates the presence of a null point at the base of the domain. Further evidence for this can found by looking at the distribution of $|\mathbf{B}|$ near the base of the domain. Figure \ref{bmag} displays this at $z=0.1$.

\subsection{Non-force-free case} 
\begin{figure}
 \resizebox{\hsize}{!}{\includegraphics{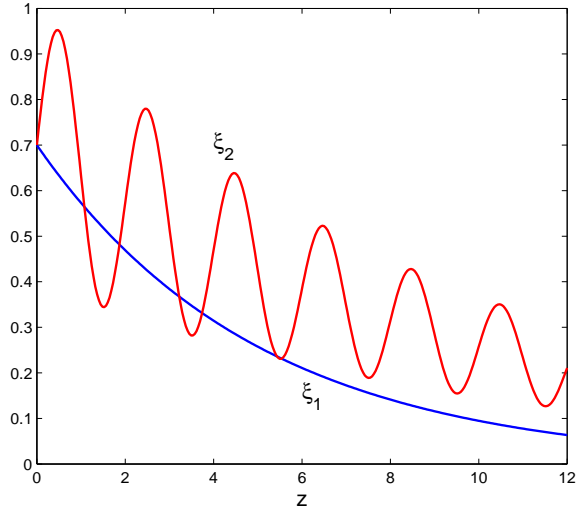}}
 \caption{The profiles of $\xi(z)$ from the cross-field current.}
\label{xi}
\end{figure}

Having demonstrated that our multigrid scheme can successfully compute linear force-free equilibria with rapidly changing field lines and null points, we now turn our attention to the linear MHS case. For this we examine two profiles for $\xi$: 
\begin{eqnarray}
\xi_1(z) &=& 0.7\exp(-0.2z), \\
\xi_2(z) &=& [0.7+0.3\sin(\pi z)]\exp(-0.1z).
\end{eqnarray}
These profiles are displayed in Figure \ref{xi}. $\xi_1(z)$ is a simple exponential decay which models the fact the magnetic field becomes more force-free as one moves from the photosphere to the corona. $\xi_2(z)$ is also exponentially decaying but has the additional complexity of a sine wave superimposed on it.
Note that both of these profiles satisfy the condition $\xi(z)<1$ to ensure that equation (\ref{scalar_pot}) is elliptic.

\subsubsection{Magnetic field and current}
\begin{figure}
 \resizebox{\hsize}{!}{\includegraphics{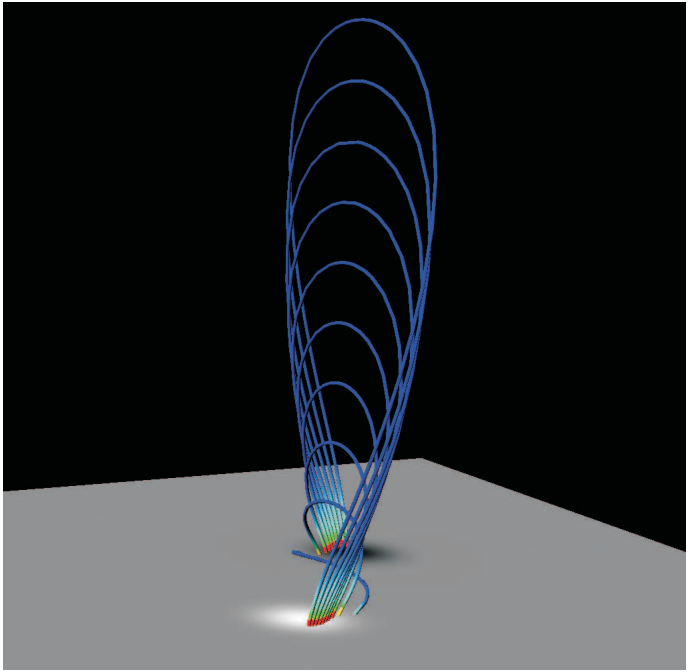}}
 \caption{Magnetic field lines for the MHS case with $\xi_1$. The field lines are coloured with $|\nabla\times\mathbf{B}|$.}
\label{profile1}
\end{figure}

\begin{figure}
 \resizebox{\hsize}{!}{\includegraphics{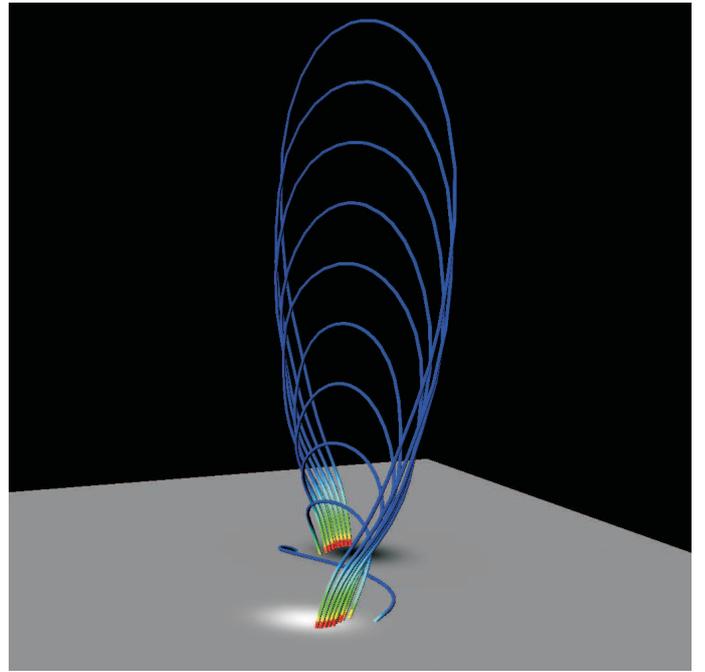}}
 \caption{Magnetic field lines for the MHS case with $\xi_2$. The field lines are coloured with $|\nabla\times\mathbf{B}|$.}
\label{profile2}
\end{figure}

The sigmoidal magnetic field for profile $\xi_1$ is displayed in Figure \ref{profile1}. That for $\xi_2$ is in Figure \ref{profile2}. In both figures, a magnetogram of $B_z$ is placed at $z=0$ and the field lines are shaded with $|\nabla\times\mathbf{B}|$.

\begin{figure}
 \resizebox{\hsize}{!}{\includegraphics{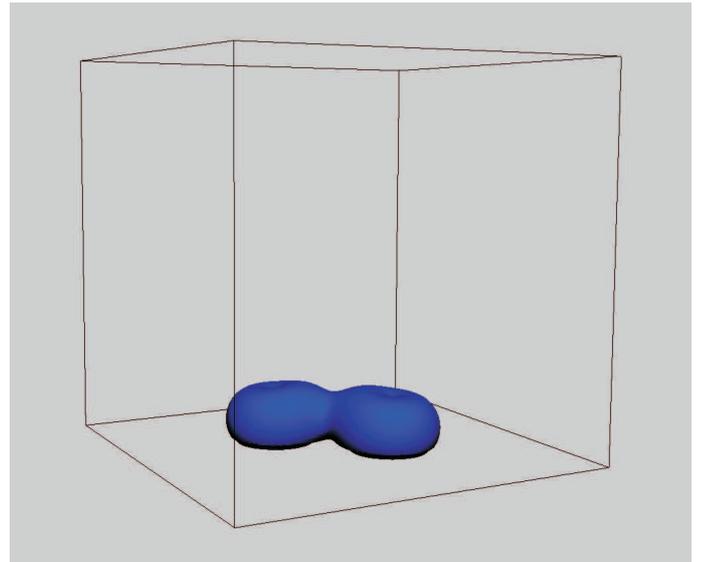}}
 \caption{An isosurface of $|\mathbf{j}| = 0.03$ for the MHS solution with $\xi_1$.}
\label{iso1}
\end{figure}

\begin{figure}
 \resizebox{\hsize}{!}{\includegraphics{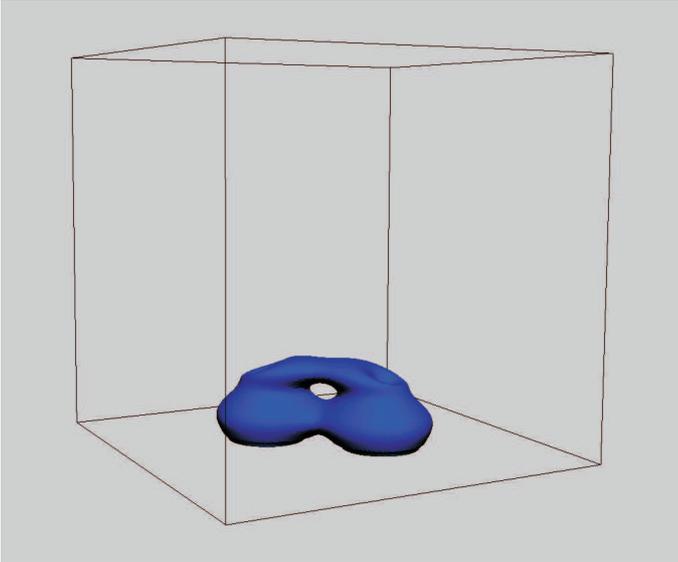}}
 \caption{An isosurface of $|\mathbf{j}| = 0.03$ for the MHS solution with $\xi_2$.}
\label{iso2}
\end{figure}

The magnetic field of $\xi_2$ is the more twisted of the two. The current is stronger at the footpoints and this is due to the initial increase in the $\xi_2$ profile. Although the magnetic fields for both profiles look similar, superficially at least, the structure of the current density is different. Figure \ref{iso1} displays an isosurface of $|\mathbf{j}|$ for the $\xi_1$ profile field. Here, the current density is concentrated at the footpoints. Figure \ref{iso2} displays the corresponding isosurface for the $\xi_2$-profile field. The oscillating $\xi_2$-profile allows for additional structure within the current density, as evidenced by an additional bridging arch in Figure \ref{iso2}. As one moves higher into the corona, the current density becomes weaker and the magnetic field becomes close to potential.  

\subsubsection{Pressure and density}
As is evident from equations (\ref{pressure}) and (\ref{density}), the plasma pressure and density are highly dependent on the magnetic field and the $\xi$-profile. Although the $\xi$-profiles are decreasing with height (on average for $\xi_2$), they are always positive. From equation (\ref{pressure}) it is clear that the effect of a positive $\xi$-profile is to introduce a pressure deficit where there is strong vertical magnetic field. Since the temperature is constant in the region near the photosphere, the geometrical depression, as measured from the pressure $p$, corresponds approximately to the Wilson depression (\cite{spruit76}). For the linear MHS equilibria of this paper, a source of size $\approx$300km produces a Wilson depression of  $\approx$100km. These values lie within the same order of magnitude as those given in \cite{spruit76}.

\begin{figure}
 \resizebox{\hsize}{!}{\includegraphics{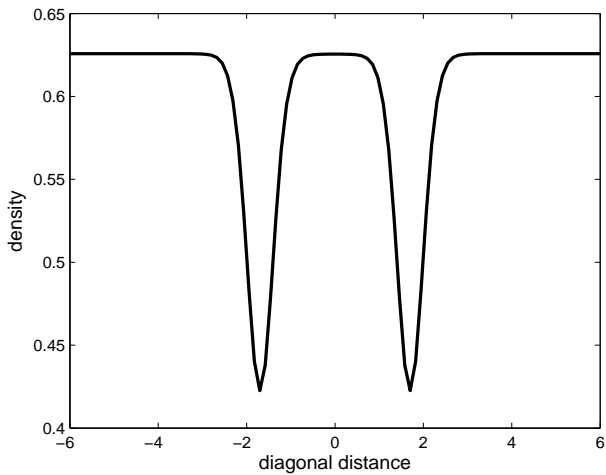}}
 \caption{The non-dimensional density along a diagonal cut through the domain at $z=0.46875$. The two troughs correspond to the locations of the magnetic footpoints. This profile is for the linear MHS equilibrium with $\xi_2$.}
\label{ddensity}
\end{figure}

Equation (\ref{density}) demonstrates that the density enhancement or depletion is due to the competition of a magnetic pressure term combined with $\xi '(z)$ and a magnetic tension term. Considering the $\xi_1$-profile first, $\xi_1 '(z) < 0$ for all $z$. Hence, the `pressure' term is negative. The `tension' term is also negative due to the fast drop in field strength of $B_z$ with height. For the $\xi_2$-profile the situation is slightly more complex as $\xi_2 '(z)$ is zero, positive and negative at different heights. This means that the `pressure' term changes sign with height. For the numerical values chosen in this paper, however, the `tension' term dominates in magnitude. This results in two density deficits over the locations of the footpoints, as shown in Figure \ref{ddensity}. If one chooses a `pressure' term that is greater in magnitude than the `tension' term, then one could produce a linear MHS model with density enhancements at different heights.

\section{Computational aspects of the implementation}

The scheme outlined in this paper reduces the problem of solving the full linear MHS equations to a finite set of scalar elliptic problems. We have implemented a multigrid method as an efficient solver of such problems. To reduce the time taken for calculations, we have made use of the massive parallelization capabilities of graphical processing units (GPUs). The development of GPUs has been driven by the computer games industry. In the last few years, however, general purpose GPU computing (GPGPU) has grown due to the development of APIs such as CUDA (for NVIDIA GPUs) and OpenCL. The use of these languages is now an important part of scientific computing. Within solar physics, GPUs have not been exploited greatly yet. This is likely to change, however, as they give scientists the capability of parallelizing their codes with hundreds or thousands of threads without having to completely redesign them.

For the problem in hand, the multigrid method involves various stages that can be treated in parallel. The relaxation steps (Gauss-Jacobi), the calculation of the residual and the interpolations between grids (reduction and prolongation) can each be treated in a parallel fashion. To perform this, one has to map the computational grid onto the GPU's grid of threads that can be run in parallel. We demonstrate this below with an example of how to perform Gauss-Jacobi iteration on a GPU. For an introduction to GPUs and programming in CUDA, we recommend \cite{sanders11}.

Subroutines that are run on the GPU are known as kernels. Below is pseudocode showing the key elements of a Gauss-Jacobi kernel used in the multigrid evaluation of equation (\ref{lower_bound}). 
\begin{verbatim}
ix = threadIdx.x + blockIdx.x*blockDim.x;
iy = threadIdx.y + blockIdx.y*blockDim.y;
off = ix + iy * blockDim.x * gridDim.x;

io = 1;
jo = DIM;

invdx2 = 1.0/(dx*dx);
invdy2 = 1.0/(dy*dy);

if ( on boundary )
{ 
	set d_P2[off] here; 
}
else {
	d_P2[off] = ( (d_P1[off+io] + d_P1[off-io])*invdx2
            + (d_P1[off+jo] + d_P1[off-jo])*invdy2
            + Bz[off] ) / (2.0*(invdx2+invdy2) );
}

\end{verbatim}
The first three lines of pseudocode link the node positions of the computational grid to that of the GPU, which consists of blocks that are divided into threads. Here we have one thread per node. In the $x$-direction, \verb0threadIdx.x0 is the thread index, \verb0blockIdx.x0 is the block index and \verb0blockDim.x0 is the number of threads per block. This combination gives the \verb0ix0 position. The \verb0iy0 position is found similarly. The third line of pseudocode calculates an offset that allows one to write all the positions on a 2D grid within a 1D array. Here, \verb0gridDim.x0 is the number of blocks in the $x$-direction. In this example, it is the same number as in the $y$-direction.

Now that a thread is linked to every node, we must be able to perform calculations involving nodes at different positions. To move to a neighbouring node in the $x$-direction, \verb0io=10 is added to or taken from \verb0off0. Similarly, to move to a neighbouring node in the $y$-direction, \verb0jo=DIM0 is added to or taken from \verb0off0. Here, \verb0DIM0 is the dimension of the grid in the $x$-direction.

With the grid set up, and the ability to move through it, the finite difference implementation of one iteration of Gauss-Jacobi relaxation is simple to implement. In the pseudocode, there is one command for points on the boundary and another for points in the interior. Values from the previous iteration are read from \verb0d_P1[]0. The new values of the Gauss-Jacobi iteration are written to \verb0d_P2[]0. Both variables begin with \verb0d0 to highlight that they are on the device (GPU) as opposed to the host (CPU).

\begin{figure*}[h!]
 \resizebox{\hsize}{!}{\includegraphics{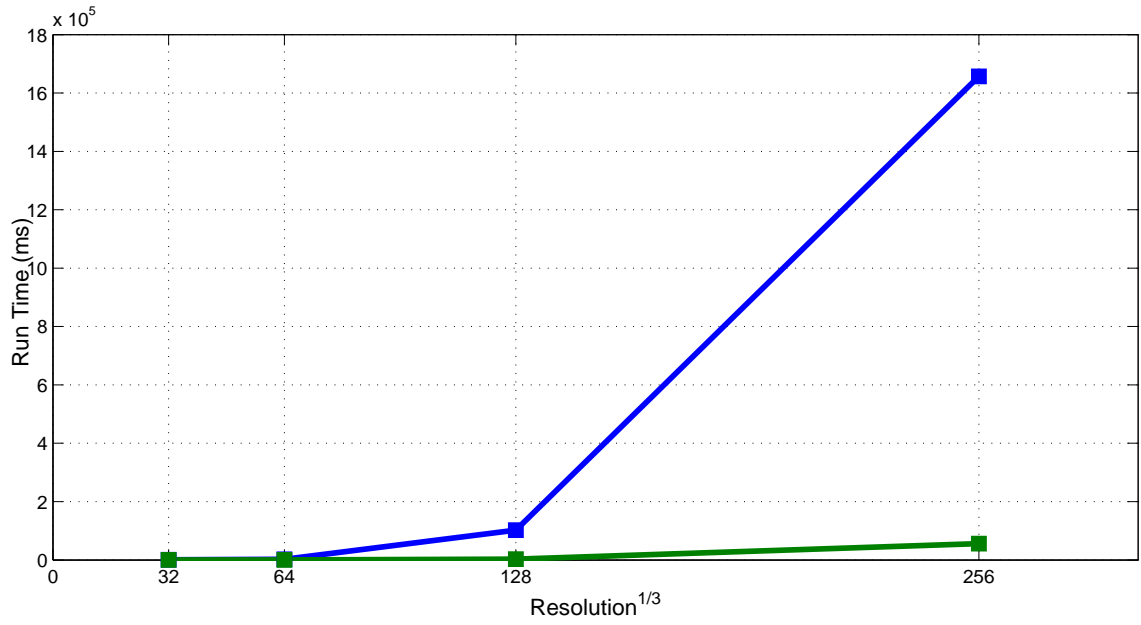}}
 \caption{The run times for the serial (blue) and parallel (green) versions of the multigrid code plotted against resolution. Each run time corresponds to the combined time for a 2D multigrid method calculation on the base of the domain and a 3D multigrid method calculation within the domain.}
\label{speedup}
\end{figure*}

This example demonstrates how one can parallelize a finite difference scheme simply. Edits to codes can be made function by function, rather than having to redesign from scratch. The purpose of this endeavour is, of course, to achieve a sizeable speed-up in the running of a code. Figure \ref{speedup} compares the run times of the serial and parallel versions of the multigrid code for different resolutions. For these calculations, we use the same parameters as for the MHS case described in the paper but set $\xi=0$. i.e. we solve the force-free case. As the resolution increases, the benefits of the parallel code run on the GPU become obvious. For resolutions of 128$^3$ and 256$^3$, impressive speed-ups of $\times$32.418 and $\times$29.632 are achieved respectively.

\section{Summary}
In this paper we calculate linear MHS equilibria with an efficient numerical scheme based on the representation of NR99. The outline algorithm of this scheme is as follows:
\begin{itemize}
\item[$\bullet$] Calculate the background equilibrium plasma plasma pressure $p_b$ and density $\rho_b$.
\item[$\bullet$] Prescribe $B_z$ on the lower boundary and find the poloidal scalar potential $P$ on this boundary by solving, using a multigrid method, the Poisson equation
\[
\frac{\partial^2P}{\partial x^2} +\frac{\partial^2P}{\partial y^2} = -B_z,
\]
with $P$ defined on the side boundaries.
\item[$\bullet$] Define the other boundary conditions for $P$, choose $\xi<1$ and $\alpha$, and solve, using a multigrid method,
\[
[1-\xi(z)]\left(\frac{\partial^2P}{\partial x^2} +\frac{\partial^2P}{\partial y^2}\right) + \frac{\partial^2P}{\partial z^2} + \alpha^2 P = 0.
\]
\item[$\bullet$] Calculate the magnetic field by applying finite differences to
\[
\mathbf{B} = \nabla\times[\nabla\times(P\hat{\mathbf{e}}_z) + \alpha P\hat{\mathbf{e}}_z].
\]
\item[$\bullet$] Calculate the pressure directly using
\[
p = p_b(z) - \xi(z)\frac{B_z^2}{2\mu_0}.
\]
\item[$\bullet$] Use finite differences to calculate the density with
\[
\rho = \rho_b(z) + \frac{1}{g}\left(\frac{{\rm d}\xi}{{\rm d}z}\frac{B_z^2}{2\mu_0} + \frac{1}{\mu_0}\xi~\mathbf{B}\cdot\nabla B_z \right).
\]
\end{itemize}
The elegance of the NR99 formulation allows for an efficient numerical solution. We demonstrate the above scheme by calculating linear force-free and linear MHS equilibria. For the force-free case, we calculate an equilibrium with three footpoints and a null point. For the linear MHS case, we consider a region with two footpoints and investigate the effects of changing $\xi$. We consider profiles that cannot be treated, in the NR99 formalism, analytically and demonstrate how these can change the structure of the current distribution within the domain. We also show that the linear MHS equilibria are consistent with the Wilson effect for the size of the regions considered.

As well as demonstrating the scheme to be accurate, we show that the calculation time can be significantly reduced through parallelization on a GPU. This is achieved function by function and speed-ups of $\times$30 are realized. This result is significant as GPUs are inexpensive and 
readily available, unlike compute clusters and supercomputers.   This fast and accurate scheme could be used for calculating equilibria in their own right or as input to MHD codes.

\begin{acknowledgements}
DM would like to thank the Royal Astronomical Society for financial support. JM acknowledges IDL support provided by STFC.
\end{acknowledgements}


\begin{thebibliography}{}

\bibitem[{Aulanier et al.} (1998)]{aulanier98}
Aulanier, G., D\'{e}moulin, P., Schmieder, B., Fang, C. \& Tang, Y.H. 1998, \solphys, 183, 369

\bibitem[{Elman et al.} (2005)]{elman05}
Elman, H.C., Silvester, D.J., Wathen, A.J. 2005, Finite Elements and Fast Iterative Solvers: with Applications in Incompressible Fluid Dynamics, Oxford University Press

\bibitem[{Henning \& Cally} (2001)]{henning01}
Henning, B.S. \& Cally, P.S. 2001, \solphys, 201, 289

\bibitem[{Hood et al.} (2012)]{hood12}
{Hood}, A.W., {Archontis}, V. \& {MacTaggart}, D. 2012, \solphys, 278, 3 

\bibitem[{{Low} (1982)}]{low82}
{Low}, B.C. 1982, \apj, 263, 952

\bibitem[{{Low} (1991)}]{low91}
{Low}, B.C. 1991, \apj, 370, 427

\bibitem[{{Low} (1992)}]{low92}
{Low}, B.C. 1992, \apj, 399, 300

\bibitem[{{MacTaggart} (2011)}]{dmac11}
{MacTaggart}, D. 2011, \aap, 531, A108

\bibitem[{McLaughlin et al.} (2012)]{mclaughlin12}
{McLaughlin}, J.A., Verth. G., Fedun, V., Erd\'{e}lyi, R. 2012, \apj, 749, 30

\bibitem[{{Neukirch} \& {Rast\"{a}tter} (1999)}]{neukirch99}
{Neukirch}, T. \& {Rast\"{a}tter}, L. 1999, \aap, 348, 1000

\bibitem[{{Petrie} \& {Neukirch} (2000)}]{petrie00}
{Petrie}, G.J.D. \& {Neukirch}, T. 2000, \aap, 356, 735

\bibitem[{{R\'{e}gnier et al.} (2005)}]{regnier05}
{R\'{e}gnier}, S., {Amari}, T. \& Canfield, R.C. 2005, \aap, 442, 345 

\bibitem[{{Sanders} \& {Kandrot} (2011)}]{sanders11}
Sanders, J. \& Kandrot, E. 2011, CUDA by Example: An Introduction to General Purpose GPU Programming, Addison-Wesley

\bibitem[{Spruit} (1976)]{spruit76}
Spruit, H.C. 1976, \solphys, 50, 269

\end{thebibliography}
\end{document}